\documentclass[12pt]{article}
\usepackage{amssymb}
\usepackage{epsfig}
\voffset= - 0.4in
\hoffset= - 0.7in         

\catcode`\@=11
\def\marginnote#1{}

\newcount\hour
\newcount\minute
\newtoks\amorpm
\hour=\time\divide\hour by60
\minute=\time{\multiply\hour by60 \global\advance\minute by-\hour}
\edef\standardtime{{\ifnum\hour<12 \global\amorpm={am}%
        \else\global\amorpm={pm}\advance\hour by-12 \fi
        \ifnum\hour=0 \hour=12 \fi
        \number\hour:\ifnum\minute<10 0\fi\number\minute\the\amorpm}}
\edef\militarytime{\number\hour:\ifnum\minute<10 0\fi\number\minute}

\def\draftlabel#1{{\@bsphack\if@filesw {\let\thepage\relax
   \xdef\@gtempa{\write\@auxout{\string
      \newlabel{#1}{{\@currentlabel}{\thepage}}}}}\@gtempa
   \if@nobreak \ifvmode\nobreak\fi\fi\fi\@esphack}
        \gdef\@eqnlabel{#1}}
\def\@eqnlabel{}
\def\@vacuum{}
\def\draftmarginnote#1{\marginpar{\raggedright\scriptsize\tt#1}}

\def\draft{\oddsidemargin -.5truein
        \def\@oddfoot{\sl preliminary draft \hfil
        \rm\thepage\hfil\sl\today\quad\militarytime}
        \let\@evenfoot\@oddfoot \overfullrule 3pt
        \let\label=\draftlabel
        \let\marginnote=\draftmarginnote
   \def\@eqnnum{(\theequation)\rlap{\kern\marginparsep\tt\@eqnlabel}%
\global\let\@eqnlabel\@vacuum}  }

\def\appname{Appendix}
\newcounter{app}
\def\theapp{\Alph{app}}
\def\app{\par
   \addvspace{4ex}
   \@afterindentfalse
  \secdef\@app\@dapp}
\def\@app[#1]#2{\ifnum \c@secnumdepth >\m@ne
        \refstepcounter{app}
        \addcontentsline{toc}{app}{\theapp
        \hspace{1em}#1}\else
      \addcontentsline{toc}{app}{ #1}\fi
   {\parindent \z@ \raggedright
    \Large \bf \appname~\theapp .
   \Large  \bf 
    #2}\nobreak
   \vskip 4ex   \noindent
\setcounter{equation}{0}
\def\theequation{\Alph{app}.\arabic{equation}}}
\def\@dapp#1{%
{\parindent \z@ \raggedright  \bf #1}\par\nobreak}
\def\l@app#1#2{\addpenalty{\@secpenalty}%
   \addvspace{1em plus\p@}%
   \begingroup
   \@tempdima 3em
     \parindent \z@ \rightskip \@pnumwidth
     \parfillskip -\@pnumwidth
     { \bf
     \leavevmode
     #1\hfil \hbox to\@pnumwidth{\hss #2}}\par
     \nobreak
   \endgroup}

\makeatletter
\newdimen\normalarrayskip            
\newdimen\minarrayskip               
\normalarrayskip\baselineskip
\minarrayskip\jot
\newif\ifold             \oldtrue            \def\new{\oldfalse}
\def\arraymode{\ifold\relax\else\displaystyle\fi}
\def\eqnumphantom{\phantom{(\theequation)}} 
\def\@arrayskip{\ifold\baselineskip\z@\lineskip\z@
     \else
     \baselineskip\minarrayskip\lineskip1\baselineskip\fi}

\def\@arrayclassz{\ifcase \@lastchclass \@acolampacol \or
\@ampacol \or \or \or \@addamp \or
   \@acolampacol \or \@firstampfalse \@acol \fi
\edef\@preamble{\@preamble
  \ifcase \@chnum
     \hfil$\relax\arraymode\@sharp$\hfil
     \or $\relax\arraymode\@sharp$\hfil
     \or \hfil$\relax\arraymode\@sharp$\fi}}

\def\@array[#1]#2{\setbox\@arstrutbox=\hbox{\vrule
     height\arraystretch \ht\strutbox
     depth\arraystretch \dp\strutbox
width\z@}\@mkpream{#2}\edef\@preamble{\halign \noexpand\@halignto
\bgroup \tabskip\z@ \@arstrut \@preamble \tabskip\z@ \cr}%
\let\@startpbox\@@startpbox \let\@endpbox\@@endpbox
  \if #1t\vtop \else \if#1b\vbox \else \vcenter \fi\fi
  \bgroup \let\par\relax
  \let\@sharp##\let\protect\relax
  \@arrayskip\@preamble}
\def\eqnarray{\stepcounter{equation}%
              \let\@currentlabel=\theequation
              \global\@eqnswtrue
              \global\@eqcnt\z@
              \tabskip\@centering              
              \let\\=\@eqncr
              $$%
            \halign to \displaywidth  \bgroup
             \eqnumphantom \@eqnsel
      \hskip\@centering                               
    $\displaystyle  \tabskip\z@ {##}$%
    &\global\@eqcnt\@ne \hskip 2\arraycolsep
         $ \displaystyle  \arraymode{##}$\hfil
    &\global\@eqcnt\tw@ \hskip 2\arraycolsep
         $\displaystyle\tabskip\z@{##}$\hfil
         \tabskip\@centering
    &{##}\tabskip\z@\cr}
\makeatother

\newfont{\hr}{msbm10}
\newfont{\ams}{msam10}

\font\numbers=cmss12
\font\upright=cmu10 scaled\magstep1
\def\stroke{\vrule height8pt width0.4pt depth-0.1pt}
\def\topfleck{\vrule height8pt width0.5pt depth-5.9pt}
\def\botfleck{\vrule height2pt width0.5pt depth0.1pt}
\def\Zmath{\vcenter{\hbox{\numbers\rlap{\rlap{Z}\kern 0.8pt\topfleck}\kern
2.2pt
                   \rlap Z\kern 6pt\botfleck\kern 1pt}}}
\def\Qmath{\vcenter{\hbox{\upright\rlap{\rlap{Q}\kern
                   3.8pt\stroke}\phantom{Q}}}}
\def\Nmath{\vcenter{\hbox{\upright\rlap{I}\kern 1.7pt N}}}
\def\Cmath{\vcenter{\hbox{\upright\rlap{\rlap{C}\kern
                   3.8pt\stroke}\phantom{C}}}}
\def\Rmath{\vcenter{\hbox{\upright\rlap{I}\kern 1.7pt R}}}
\def\Z{\ifmmode\Zmath\else$\Zmath$\fi}
\def\Q{\ifmmode\Qmath\else$\Qmath$\fi}
\def\N{\ifmmode\Nmath\else$\Nmath$\fi}
\def\C{\ifmmode\Cmath\else$\Cmath$\fi}
\def\R{\ifmmode\Rmath\else$\Rmath$\fi}

\def\d{\partial}

\def\bea{\begin{eqnarray}}
\def\eea{\end{eqnarray}}

\def\beq{\begin{equation}}
\def\eeq{\end{equation}}
\def\ba{\beq\new\begin{array}{c}}
\def\ea{\end{array}\eeq}
\def\be{\ba}
\def\ee{\ea}
\def\F{{\cal F}}

\def\stackreb#1#2{\mathrel{\mathop{#2}\limits_{#1}}}

\def\res{{\rm res}}

\def\Re{{\rm Re}}

\def\half{{\textstyle{1\over2}}}
\def\ha{{1\over 2}}
\def\N2{${\cal N}=2$}
\def\4N{${\cal N}=4$}
\def\1N{${\cal N}=1$}
\def\1N*{${\cal N}=1^*$}

\def\beq{\begin{equation}}
\def\eeq{\end{equation}}
\def\ba{\beq\new\begin{array}{c}}
\def\ea{\end{array}\eeq}
\def\be{\ba}
\def\ee{\ea}
\newcommand{\rf}[1]{(\ref{#1})}

\begin{document}


\begin{flushright}
FIAN/TD-26/08\\
ITEP/TH-44/08
\end{flushright}
\vspace{0.5 cm}

\renewcommand{\thefootnote}{\fnsymbol{footnote}}
\begin{center}
{\Large\bf
Non Abelian gauge theories, prepotentials\\
\vspace{0.4 cm}
and Abelian differentials\footnote{Based
on talks, given at {\em Workshop on combinatorics of moduli spaces,
Hurwitz numbers, and cluster algebras}, Moscow;
{\em Abel Symposium 2008}, Troms{\o} and {\em Geometry and integrability in mathematical
physics 08}, Luminy.}
}\\
\vspace{1.0 cm}
{\large A.~Marshakov}\\
\vspace{0.6 cm}
{\em
Theory Department, P.N.Lebedev Physics Institute,\\
Institute of Theoretical and Experimental Physics,\\ Moscow, Russia
}\\
\vspace{0.3 cm}
{e-mail:\ \ mars@lpi.ru,\ \ mars@itep.ru}
\end{center}
\begin{quotation}
\noindent
I discuss particular solutions of the integrable systems, starting from well-known
dispersionless KdV and Toda hierarchies, which define in most straightforward way the
generating functions for the Gromov-Witten classes in terms of the rational complex
curve. On the ``mirror'' side these generating functions can be
identified with the simplest prepotentials of complex manifolds, and I present
few more exactly calculable examples of them. For the higher genus curves, corresponding
in this context to the non Abelian gauge theories via the topological gauge/string duality,
similar solutions are constructed using extended basis of Abelian differentials, generally
with extra singularities at the branching points of the curve.
\end{quotation}

\renewcommand{\thefootnote}{\arabic{footnote}}
\setcounter{section}{0}
\setcounter{footnote}{0}
\setcounter{equation}0

\newpage
\section{Introduction}

Integrable differential equations appear in different branches of modern mathematical physics,
but in the last years they attracted a lot of attention, due to the study of partition function
in the simplest models of string theory and some quantum gauge theories.
The generating functions for the
correlators, which can be symbolically written as
\be
\label{pafu}
\log\tau({\bf t}) = \langle\exp\sum_k t_k\sigma_k\rangle_{\rm string}
\ee
are initially defined by
summing the perturbation series and even the instanton expansions in world-sheet formulations
of a string model or in a relatively simple quantum field theory. It turns out, however, that
all essential information about the results of this summation is typically hidden
in a system rather simple (though generally nonlinear) differential equations for $\log\tau({\bf t})$
and its derivatives.

In a certain sense the theoretical physicists are lucky:
both for the toy-models and ``physical'' multi-dimensional target-space theories one gets
rather well-known and widely used for the applied problems of mathematical physics integrable
systems in $1+1$ and $2+1$ dimensions. Moreover, quite often we are interested even in the solutions
with the {\em finite} number of degrees of freedom - the so called moduli of the theory (corresponding
to the
finite-dimensional set of the {\em primary} operators and corresponding time-parameters, sometimes
also called as ``small phase space''). The geometry of these solutions involve mostly the complex
curves and their Jacobians, so that the Abel map plays the role of an integrating change of the
variables.

Being closely related to the algebro-geometric solutions of the KP and Toda-type
integrable systems, the {\em string solutions} are nevertheless absolutely new. In order
to demonstrate this the best thing is to start from two well-known examples.

{\bf 1}: The famous KdV equation $u_t + uu_x + u_{xxx} = 0$ is trivially satisfied by the function
$u = {x\over t}$, which is nothing but a linear potential, ``slowly falling down'' in time. This rather
trivial solution from the point of view of KdV equation itself corresponds to the
pure topological gravity (simplest topological string model) or the Gromov-Witten
theory of a point \cite{WDW,KSch,Konts,gkm}, with the partition function (known as Kontsevich
model)
\be
\label{utko}
\log\tau = {\sf F}_K(x,t) + \ldots = {x^3\over 6t} + \ldots \stackreb{t\to t+1}{=}\
{x^3\over 3!}\langle{\bf 1}{\bf 1}{\bf 1}\rangle + \ldots
\\
u = {\d^2\log\tau\over\d x^2}
\ee
producing the intersection numbers,
\be
\label{innu}
\langle\sigma_{k_1}\ldots\sigma_{k_n}\rangle = \int_{{\overline M}_{g,n}}\prod_{i=1}^n\psi_i^{k_i}
\ee
defined as integrals over compactified moduli spaces ${\overline M}_{g,n}$ of genus $g$ complex curves
with $n$ punctures (the world-sheets for the $n$-point correlators in string theory),
where $\psi_i = c_1(\mathbb{L}_i)$ is the first Chern class
of canonical line bundle on ${\overline M}_{g,n}$ with
the fiber $T^\ast_{\Sigma_{g,n}} (P_i)$ at $i$-th marked point $P_i$.
Literally in \rf{utko} the only intersection number $\langle{\bf 1}{\bf 1}{\bf 1}\rangle =1$ is
written down,
corresponding to the trivial integral over ${\overline M}_{0,3}={\rm point}$. In order to get the
rest of the numbers \rf{innu},
one needs to solve the full KdV hierarchy (to switch on higher flows in rest of the
variables $t_1=x,t_3=t,t_5,t_7\ldots$ of the hierarchy) for the initial $u=x/t$
\be
\label{tako}
\log\tau = \sum_{\{k_i\}\geq 0} {t_{2k_1+1}\ldots t_{2k_n+1}\over n!}
\langle\sigma_{k_1}\ldots\sigma_{k_n}\rangle\hbar^{2g-2}
\ee
In physical language one gets by \rf{tako} the generating function for the correlators
of the {\em gravitational descendants} $\sigma_k \equiv \sigma_k({\bf 1})$, corresponding
to the multiplication of the primary operators by $k$-th powers of the Chern classes.
The ``target-space'' part of the theory of a point is trivial and
contains the only primary unity operator ${\bf 1}$.
For convenience, the string coupling $\hbar$ is
introduced in \rf{tako}, with the weight, fixed by selection rule $\sum k_i = 3g-3+n$.
For example, the contribution explicitly presented in \rf{utko} is weighted by $\hbar^{-2}$,
since it comes from the moduli space ${\overline M}_{0,3}$ with $g=0$.

This extra parameter is needed, since of special interest is the quasiclassical limit $\hbar\to 0$
of the generating function, which behaves then as
\be
\label{quta}
\log\tau ({\bf t})\ \stackreb{\hbar\to 0}{\sim}\ {1\over\hbar^2}\F ({\bf t}) + O(\hbar^0)
\ee
with $\F$ often called as prepotential. The quasiclassical part of the expansion \rf{tako}
is described in terms of dispersionless limit for the KdV equation with the Lax {\em function}
$W = z^2 - u$ (the Lax operator after the substitution
$\d/\d x\to \hbar\d/\d x\to z$), or explicitly by the formulas
\be
\label{resko}
x = \res_{\infty} W^{-1/2}zdW \sim u
\\
{\d\F\over \d x} = \res_{\infty} W^{1/2}zdW \sim u^2
\ee
so that
\be
\label{fko}
\F\ \stackreb{\hbar\to 0}{=}\ \hbar^2\log\tau = {\sf F}_K(x) \sim x^3 + \ldots
\ee
Formulas like \rf{resko} define the prepotential \rf{fko}, the quasiclassical part of \rf{tako},
also if all gravitational descendants are added, as degenerate
prepotential of an almost trivial complex manifold - the target-space rational curve $W = z^2 - u$.
Generally one has many variables, the residues should be replaced by the period integrals
$\res\to\oint$ over all nontrivial cycles, and integrability of the equations like \rf{resko}
is guaranteed by the Riemann bilinear identities.

The full partition function $\tau ({\bf t})$ can be also
restored \cite{FKNDVV} as a solution to the Virasoro
constraints
\be
L_n\tau = 0,\ \ \  n\geq -1,
\\
L_n = {1\over 2}\sum kt_k{\d\over\d t_{k+2n}} + {1\over 4}\sum _{a+b=2n}
{\partial ^2\over \partial t_a\partial t_b}+
\\
+ \delta _{n+1,0}{t^2_1\over 4} + {\delta _{n,0}\over 16}
\ee
being an infinite set of the {\em linear} differential equations.

{\bf 2}: Consider now the second simplest example of the
Toda chain \cite{Toda} (in dispersionless limit):
\be
\label{dito}
{\d^2\F\over\d t_1^2} =
\exp {\d^2\F\over\d a^2}
\ee
The ``stringy solution''
\be
\label{fto}
\F = \half a^2t_1 + e^{t_1}
\ee
(again for the prepotential
$\F \stackreb{\hbar\to 0}{=}\ \hbar^2\log\tau$)
describes system of particles with the co-ordinates $a^D = \d\F/\d a = at_1$ moving
with constant ${\rm velocity} = {\rm number} = a$ in time $t_1$, being the first time of the
Toda chain hierarchy.
These two parameters $(a,t_1)$ replace here the only ``target-space'' parameter $x$ of the KdV
hierarchy, since the primary operators here - instead of the only ${\bf 1}$ in the
KdV example - correspond to the cohomologies of $\mathbb{P}^1$:
$a \leftrightarrow {\bf 1}\in H^0(\mathbb{P}^1)$ and
$t_1 \leftrightarrow \varpi\in H^2(\mathbb{P}^1)$. The truncated generation function
$\F \sim \langle \exp\left({a{\bf 1}+t_1\varpi}\right)\rangle$ gives rise to the deformation
of the multiplication in cohomology ring:
$\varpi\cdot\varpi \simeq e^{t_1}{\bf 1}$, corresponding to the only nontrivial relation
in the operator algebra of the target-space primary operators.

To restore the dependence upon the gravitational descendants $t_{k+1} \leftrightarrow \sigma_k(\varpi)$,
$T_n \leftrightarrow \sigma_n(\bf{1})$, (in these notations $a\equiv -T_0$) one has to solve
the Toda chain hierarchy, with the initial condition, corresponding to \rf{fto}. Quasiclassically,
for the prepotential
\be
\F = {a^2 t_1\over 2} + e^{t_1} \Rightarrow \F({\bf t},a) \Rightarrow \F({\bf t},{\bf T})
\ee
it can be done in two steps (certainly, both the half-truncated generating function
$\F({\bf t},a)$ and full $\F({\bf t},{\bf T})$
still satisfy the first Toda equation \rf{dito}).
Solving Toda chain hierarchy in ${\bf t}$-variables gives rise to a (half-truncated)
Gromov-Witten theory of complex projective line $\mathbb{P}^1$
\be
\label{tautr}
\log\tau = \sum_{\{k_i\},d\geq 0} {t_{k_1}\ldots t_{k_n}\over n!}
\langle\sigma_{k_1}(\varpi)\ldots\sigma_{k_n}(\varpi)\rangle\hbar^{2g-2}q^{d}
\ee
where $\sigma_k(\varpi)$ are the descendants
of the (generally complexified) K{\"a}hler class $\varpi$, (to distinguish them
from the descendants $\sigma_k({\bf 1})$ of the unity operator) and
the correlators are now identified
\be
\label{nup1}
\langle\sigma_{k_1}\ldots\sigma_{k_n}\rangle = \int_{{\overline M}_{g,n}(\mathbb{P}^1,d)}
\prod_{i=1}^n\psi_i^{k_i}{\rm ev}^\ast_i(\varpi)
\ee
with the integrals taken over the moduli spaces
${\overline M}_{g,n}(\mathbb{P}^1,d)$ of stable maps of degree $d$, and
${\rm ev}_i: {\overline M}_{g,n}(\mathbb{P}^1,d)\mapsto \mathbb{P}^1$ is
evaluation map at the $i$-th marked point. The extra parameter $q$ in \rf{nup1}, counting
degree of the maps, or the number of instantons, can be absorbed by shift of
$t_1\to t_1-\log q$.

The quasiclassical part of the generating function \rf{tautr} is again described by a prepotential
in a dual picture, sometimes also called as the ``Landau-Ginzburg approach''. The
Landau-Ginzburg superpotential can be now chosen as a function on cylinder
\be
\label{dtcu}
z = v + \Lambda\left(w+{1\over w}\right)
\ee
which has also an obvious sense of the Lax function of the dispersionless Toda chain (the
r.h.s. of \rf{dtcu} represents the three-diagonal Lax matrix of Toda chain in terms
of powers of spectral parameter $w$).
Equation \rf{dtcu} can be viewed as describing a rational curve (a cylinder), embedded into
$(z,w)\subset\mathbb{C}\times\mathbb{C}^\ast$, and it can be considered as a particular oversimplified
example of the $N_c$-periodic Toda chain curves family
\be
\label{Todacu}
\Lambda^{N_c}\left(w+{1\over w}\right) = P_{N_c}(z) = \prod_{i=1}^{N_c} (z-v_i)
\ee
The topological type-A string theory on $\mathbb{P}^1$ is in this way dual to the
$N_c=1$ oversimplified Abelian \N2 supersymmetric gauge theory \cite{LMN}. Solving the
dispersionless Toda chain, corresponding to \rf{dtcu}, is therefore an intermediate step
towards understanding the geometry of the extended non Abelian \N2 supersymmetric gauge theory
with the $U(N_c)$ gauge group, being associated in the Seiberg-Witten context \cite{gkmmm} with the
families of the curves \rf{Todacu}.

\section{Topological solution to dispersionless Toda hierarchy}

The solution for the half-truncated dispersionless Toda hierarchy for $\F({\bf t},a)$ at
$T_n=\delta_{n,1}$, or switched off gravitational descendants of unity $\{\sigma_k({\bf 1})\}$
is given in terms of the rational curve \rf{dtcu} (or the dispersionless Toda Lax operator),
endowed with
\be
\label{Sz}
S\stackreb{z\to\infty}{=} -2z(\log z -1) + \sum_{k>0} t_kz^k +2a\log z -
{\d\F\over\d a} - 2\sum_{k>0} {1\over kz^k}{\d\F\over\d t_k}
\ee
odd under the involution $w\leftrightarrow{1\over w}$,
which has the sense of the logarithm $S\sim\log\Psi$ of $\Psi$-function, solving the
auxiliary linear problem.
As always for the integrable systems, solution for the dynamical variables themselves
comes from reconstructing $\Psi$ or $S$.
In terms
of the global variable $w$ one immediately writes \cite{MN}
\be
\label{Sw}
S = -2\left(z\log w + \Lambda(\log\Lambda-1)\left(w-{1\over w}\right)\right)
+ \sum_{k>0} t_k\Omega_k(w) +2a\log w
\ee
fixed by asymptotic $w\stackreb{z\to\infty}{\sim}z$ and being odd under
$w\leftrightarrow{1\over w}$. Here $\Omega_k  = z(w)^k_+-z(w)^k_-$,
where $\pm$ stand for the strictly positive and negative parts of the Laurent polynomials
(powers of $z$ \rf{dtcu}) in $w$, e.g.
$\Omega_1(w) = \Lambda\left(w-{1\over w}\right)$, $\Omega_2(w) = \Lambda^2\left(w^2-{1\over w^2}\right)
+ 2\Lambda v\left(w-{1\over w}\right)$, etc.

Expressions for $v$, $\Lambda$, $\F$ as functions of $a$ and all times ${\bf t}$ are found from
the conditions
\be
\label{eqscu}
\left.{dS\over d\log w}\right|_{dz =0} = 0
\ee
imposed at the zeroes of the differential $dS$ coinciding with the zeroes of $dz$,
i.e. at the ramification points \cite{KriW}. These are two algebraic
equations, to be solved for the coefficients $v=v(a;{\bf t})$ and $\Lambda=\Lambda(a;{\bf t})$
of the curve \rf{dtcu}.

\bigskip
\noindent
{\bf Small phase space}.
If, for example, $t_k=0$ for $k>1$, it gives
\be
\label{t01}
v=a, \ \ \ \Lambda^2=e^{t_1}
\ee
and from the ``regular tail'' of the expansion \rf{Sz} one reads off the
prepotential \rf{fto} on the small phase space.

One can interpret this as a particular degenerate case of a
general definition of prepotential of a complex curve $\Sigma$, endowed with two meromorphic
differentials with the fixed periods \cite{KriW}, or with the generating Seiberg-Witten
one-form $dS_{SW}$. The
variables are generally introduced via the period integrals
\be
\label{aper}
a_i={1\over 4\pi i}\oint_{A_i}dS_{SW}
\ee
over the
chosen half of the cycles from $H_1(\Sigma)$,
and the gradients of prepotential $\F$ determined
\be
\label{aD}
a^D_i=\oint_{B_i}dS_{SW} = {\partial {\cal F}\over\partial a_i}
\ee
by the period integrals over the dual cycles. The definition \rf{aD} is consistent
due to condition
\be
\label{prepsw}
{\d a^D_i\over\d a_j} = T_{ij} = {\d^2{\cal F}\over\d a_i\d a_j}
\ee
where the symmetricity of the r.h.s. (or of the second derivatives
of the prepotential)
is guaranteed by symmetricity of the period matrix of $\Sigma$.
Equality \rf{prepsw} follows from \rf{aD} and is implied by the fact that
variation of $dS_{SW}$ w.r.t. moduli is holomorphic, that is a direct analog
of the property \rf{eqscu}.

The above solution for dispersionless Toda with only nonvanishing $a$ and $t_1$ is
just a degenerate version for this construction, where the nontrivial curve is replaced by a
cylinder. Then
\be
\label{SLeg}
S = -2\left(z\log w + \Lambda(\log\Lambda-1)\left(w-{1\over w}\right)\right)+
t_1\Lambda\left(w-{1\over w}\right) + 2a\log w =
\\
\stackreb{\rf{t01}}{=}\ -2z\log w + 2\Lambda \left(w-{1\over w}\right) + 2a\log w
\\
dS_{SW} = 2z{dw\over w}
\ee
so that
\be
\label{ares}
{1\over 4\pi i}\oint_{A_i}dS_{SW} = {1\over 2\pi i}\oint_A z{dw\over w} =
\res_{z=\infty}\ z{dw\over w} = a
\ee
($dz$ and ${dw\over w}$ are these
two meromorphic differentials with the fixed periods), and
\be
\label{aDS0}
{\d\F\over\d a} \sim \int_B z{dw\over w} \sim [S]_0 = at_1
\ee
where the role of the regularized degenerate infinite $B$-period is played by the constant part of the
function \rf{Sz}.

\bigskip
\noindent
{\bf Higher flows}. To add the higher flows, one have to introduce the generalized periods, or
just the coefficients of the expansion \rf{Sz}, which can be denoted as
\be
\label{tP}
t_k = {1\over k}\ \res_{P_+} z^{-k}dS =
- {1\over k}\ \res_{P_-} z^{-k}dS,\ \ \ k>0
\ee
and
\be
\label{tPd}
{\d\F\over \d t_k} =  \ha \res_{P_+} z^{k}dS
= - \ha \res_{P_-} z^{k}dS,\ \ \ k>0
\ee
where $z(P_+)=z(P_-)=\infty$ and these are two infinities of \rf{dtcu}, exchanged by the involution
$w\leftrightarrow{1\over w}$. Equations \rf{eqscu} remain the same, but they cannot
be solved explicitly in general.

Already adding only nonvanishing $t_2$ their solutions \cite{MN}
\be
\label{t012}
v=a -{1\over 2t_2}{\bf L}\left(-4t_2^2e^{t_1+2t_2a}\right)
\\
\log\Lambda^2 = t_1+2t_2a - {\bf L}\left(-4t_2^2e^{t_1+2t_2a}\right)
\ee
are expressed only in terms of the Lambert function ${\bf L}(x) e^{{\bf L}(x)}= x$.
This is nothing, but the asymptotic of the generation function for the Hurwitz numbers
\be
\label{hunu}
H_{g,d} = \langle\sigma_1(\varpi)^{2g+2d-2}\rangle_{g,d}
\ee
each of them having a meaning of
the number of genus $g$, $d$-sheeted covers of $\mathbb{P}^1$, with a fixed general
branch divisor of degree $d\cdot\chi(\mathbb{P}^1)-\chi(\Sigma_g) = 2d+2g-2$,
as follows from the Riemann-Hurwitz formula.

Indeed, one gets from \rf{tautr}, \rf{hunu}
\be
\label{fhu}
\F (a=0,t_1,t_2=\half,0,\ldots) = \sum_{d>0}{H_{d,0}\over (2d-2)!}e^{dt_1}
\ee
From our solution \rf{t012} it follows
\be
\left.{\d^2\F\over\d t_1^2}\right|_{a=0,t_2=1/2} = \Lambda^2 = - {\bf L}\left(-e^{t_1}\right)
\ee
which produces exactly $H_{d,0}={(2d-2)!\over d!} d^{d-3}$ since the Lambert function
has an expansion
\be
{\bf L}(t) = \sum_{n=1}^{\infty} {(-n)^{n-1} t^{n}\over n!}
= t - t^2 + {3\over 2} t^{3} - {8 \over 3} t^{4} + \ldots
\ee
giving rise to the desired result.

\section{Quasiclassics of Nekrasov partition function}

The tau-function \rf{tautr} can be in fact defined beyond quasiclassical (corresponding
to $g=0$ Gromov-Witten potential) theory \cite{Nek,LMN}. The definition can be written as sum over
partitions: the sets of integers ${\bf k} = ( k_1  \geq
k_2 \geq \ldots \geq k_{\ell_{\bf k}}=0 \geq 0 \ldots)$
\be
\label{zuone}
\tau( a, {\bf t}) =
\sum_{{\bf k}}
{{\bf m}_{\bf k}^2\over{( - {\hbar}^2)^{| {\bf k} |}}} \
e^{{1\over {\hbar}^{2}} \sum_{k>0}
{t_{k}\over k+1}{{\rm ch}_{k+1} ( a, {\bf k},\hbar)}}\sim
\exp\left({1\over\hbar^2}\F( a, {\bf t}) + \ldots\right)
\ee
weighted with the squared Plancherel measure
\be
\label{planch}
{\bf m}_{\bf k}
=  {\prod_{1\leq i<j \leq \ell_{\bf k}}(k_i-k_j+j-i)\over
\prod_{i=1}^{\ell_{\bf k}}(\ell_{\bf k}+k_i-i)!}
\sim\prod_{i<j}{k_i-k_j+j-i\over j-i}
\ee
and coupled to the Toda times by the Chern polynomials
\be
\left( e^{{\hbar}u\over 2} -
e^{-{\hbar u \over 2}} \right)
\sum_{i=1}^{\infty} e^{ u ( a + \hbar({\half} - i +k_{i}))}=
  \sum_{l=0}^{\infty}\
{u^l\over l!}\ {\rm ch}_{l}( a, {\bf k};\hbar)
\ee
or
\be
{\rm ch}_{0}(a, {\bf k})  = 1,\ \ {\rm ch}_{1}( a, {\bf k} ) = a, \\
{\rm ch}_{2} ( a, {\bf k}) = a^2  +
2{\hbar}^{2} \vert {\bf k} \vert
\\
{\rm ch}_{3}( a, {\bf k}) = a^{3} + 6{\hbar}^{2} a \vert {\bf k} \vert +
3{\hbar}^{3} \sum_{i}
k_{i} ( k_{i} + 1 - 2i )
\\
\ldots
\ee
The expression in the r.h.s. of the last one has an easily recognizable ingredient
(see e.g. \cite{OkTo})
\be
\label{cohu}
\sum_{i} k_{i} ( k_{i} + 1 - 2i ) = \sum_{i} \left((k_i-i+\half)^2-(-i+\half)^2\right)
\ee
from the combinatorics of Hurwitz numbers - the class of a transposition, and it is exactly
the element, whose coupling to $t_2=\half$ in \rf{fhu} ensures appearance of the asymptotic
of Hurwitz numbers via the expansion of the Lambert function.

Topological gauge string duality reinterprets the
sum over partitions in the expression for the exponentiated full Gromov-Witten potential
\rf{zuone} as summing over all
instantons in the deformed four-dimensional  \N2 supersymmetric gauge theory \cite{LMN}.
Expression \rf{zuone} is a particular example of the Nekrasov partition function \cite{Nek}
for the $N_c=1$ or deformed $U(1)$ gauge theory.

Formula \rf{zuone} also states, that quasiclassics $\hbar\to 0$ of the Nekrasov partition function
coincides with the genus zero Gromov-Witten potential or the Seiberg-Witten prepotential of
the extended $U(1)$ theory. This equivalence leads, in particular, to the strange phenomenon -
effective actions in four-dimensional supersymmetric gauge theories satisfy {\em the same}
differential equations as generating functions for the correlators in
topological strings!

Quasiclassical contribution into \rf{zuone} can be found as a
solution to the extremum problem for the functional
\be
\label{functnl}
\F = \half\int dx f''(x)\sum_{k>0}t_k{x^{k+1}\over k+1}-
\half\int_{x_1>x_2} dx_1 dx_2 f''(x_1)f''(x_2) F(x_1-x_2) +
\\
+ a^D\left(a-\half\int dx\ xf''(x)\right) + \sigma\left(1-\half\int dx\ f''(x)\right)
\ee
whose form is derived from the integral representation of the Chern polynomials
\be
\label{chernf}
{\rm ch}_{l}(a, {\bf k}) = \half \int \ dx \ f_{\bf k}''(x) x^l
\sim \sum_{i=1}^{\infty}
\left( (a + {\hbar} ( k_{i} - i +1 ))^l - (a +{\hbar}
( k_{i} - i  ))^l\right)
\ee
and the Plancherel measure
\be
\label{plg}
{\bf m}_{\bf k}^2 \sim \prod_{i,j}\left(k_i-k_j+j-i\right)
= \exp\sum_{i,j}\log\left(k_i-k_j+j-i\right)\sim
\\
\sim \exp\left(-{1\over 2\hbar^2}\int_{x_1>x_2} dx_1 dx_2 f_{\bf k}''(x_1)f_{\bf k}''(x_2)
\gamma(x_1-x_2;\hbar)\right)\ \sim
\\
\stackreb{\hbar\to 0}{\sim}\ \exp\left(-{1\over 2\hbar^2}\int_{x_1>x_2} dx_1 dx_2
f_{\bf k}''(x_1)f_{\bf k}''(x_2)
F(x_1-x_2)\right)
\ee
via the second derivative of the shape function \cite{NO}
\be
\label{fpp}
f_{\bf k}''(x) \sim
2\sum_{i=1}^\infty \left(\delta(x-a-{\hbar}({k}_{i}-i +1)) -
\delta(x-a -{\hbar}({k}_{i}-i))\right)
\ee
for partitions.
In \rf{plg} the kernel $\gamma(x;\hbar)$ satisfies the second order difference equation
\be
\label{gammaeq}
\gamma\left(x+\hbar\right) + \gamma\left(x-\hbar\right)-2\gamma\left(x\right) = \hbar^2\log x
\ee
and for $\hbar\to 0$ can be replaced in the main order by
\be
\label{gamf}
\gamma(x;\hbar)\ \stackreb{\hbar\to 0}{\Rightarrow}\ F(x) = {x^2\over 2}\left(\log x - {3\over 2}\right)
\ee
or just $F''(x) = \log x$, known as perturbative prepotential for the four-dimensional
\N2 supersymmetric gauge theory.

The shape function
\be
\label{shfu}
f_{\bf k}(x) = |x-a|+\Delta f_{Y_{\bf k}}(x) \sim \sum_{i=1}^\infty \left(
|x-a-{\hbar}({k}_{i}-i +1)|-|x-a -{\hbar}({k}_{i}-i)|\right)
\ee
literally corresponds to the shape of the Young diagram $Y_{\bf k}$ of partition ${\bf k}$, put
into the right angle $|x-a|$ whose vertex is located at $x=a$ of the $x$-axis. The functional
\rf{functnl} should be computed on the extremal partition ${\bf k}_\ast$, corresponsing to some
``large'' partition $Y_{{\bf k}_\ast}$ with the shape
function $f_{{\bf k}_\ast}(x)\equiv f(x)$, to be found as solution for the extremal equation for
the functional \rf{functnl}. Two last terms in the r.h.s. of \rf{functnl} reflect added with the Lagrange
multipliers constraints for the shape function, following from \rf{shfu}:
$f'_{\bf k}(x^+)-f'_{\bf k}(x^-)=2$ corresponding to approaching of the right angle by the
shape function $f_{\bf k}(x^\pm)=|x^\pm-a|$ at certain points $x^\pm$, and location of the vertex
of the Young diagram and the right angle at $a = \half\int dx\ xf_{\bf k}''(x)$.

Extremizing the functional \rf{functnl}, one gets for
$S(z) = {d\over dz}{\delta\F\over\delta f''(z)}$, or
\be
\label{Sfun}
S(z) = \sum_{k>0}t_k z^k - \int dx f''(x)(z-x)\left(\log(z-x)-1\right)-a^D
\ee
that its real part
\be
\label{Scut}
\Re\ S(z) = \half\left(S(z+i0)+S(z-i0)\right) =  0,\ \ \ \ z\in {\bf I}
\ee
vanishes on the segment ${\bf I}$, where $\Delta f(x) \neq 0$: $x^-<x<x^+$. The asymptotic of \rf{Sfun}
at $z\to\infty$ coincides with \rf{Sz}, and to construct such function, satisfying \rf{Scut}
one takes the double cover $y^2=(z-x^+)(z-x^-)$ of the $z$-plane or the cylinder \rf{dtcu}
with $x^\pm = v\pm 2\Lambda$, and writes literally the odd under exchange of the two $z$-sheets
expression \rf{Sw}, which automatically obeys \rf{Scut}.

The extremal shape function is found from the \rf{Sw} as
$f'(x) \sim {\rm jump}\ \left({dS\over dx}\right)$, as follows from the integral representation
\rf{Sfun}. It has been found in \cite{MN}, for example, that if one adds nonvanishing $t_2 \neq 0$
to the small phase space, the extremal shape function equals
\be
f^{\prime}(x) = {2\over {\pi}} \left( {\rm arcsin}
\left( {x - v \over 2{\Lambda}} \right) + 2t_{2}
\sqrt{4 {\Lambda}^{2}-(x-v)^2} \right), \\
v - 2 \Lambda \leq x \leq v + 2 \Lambda
\ee
i.e. the Vershik-Kerov arcsin law is deformed by the Wigner semicircle and
``renormalization'' $v=a\to v(a;{\bf t})$ and $\Lambda=e^{t_1/2}\to\Lambda(a;{\bf t})$ of
the parameters of the curve, solving equations \rf{eqscu}.

\section{Full quasiclassical Gromov-Witten potential of $\mathbb{P}^1$}

To restore ${\bf T}$-dependence in the partition function
$\tau(a,{\bf t}) \rightarrow \tau(a,{\bf t},{\bf T})$ or to switch on the descendants of
unity $\{ \sigma_k({\bf 1})\}$ for $k>0$ one has to solve
the Virasoro constraints
\be
L_n ({\bf t},{\bf T};\d_{\bf t},\d_{\bf T};\d^2_{\bf t})\tau(a,{\bf t},{\bf T})=0,\ \ \ n\geq -1
\ee
with the initial condition $\tau(a,{\bf t}) = \left.\tau(a,{\bf t},{\bf T})\right|_{T_n=\delta_{n,1}}$,
see \cite{EHY,Getz,Giv,OP,Dub}. Quasiclassically, solution to these Virasoro constraints, producing
the full genus zero Gromov-Witten potential $\F(a,{\bf t},{\bf T})$ is described \cite{AMextT}
by the following
generalization of the formula \rf{Sz}
\be
\label{Szfu}
S(z)\ \stackreb{z\to\infty}{=}\ \sum_{k>0}t_kz^k
 - 2\sum_{n>0}T_nz^n(\log z-c_n)
  +
 \\
 + 2a\log z - {\d\F\over\d a}
- 2\sum_{k>0} {1\over kz^k}{\d\F\over\d t_k}
\ee
($c_k = \sum_{i=1}^k{1\over i}$ are harmonic numbers), which
defines $\F$ when constructed globally on \rf{dtcu}, odd under the involution
$w\leftrightarrow{1\over w}$.

To do this, one needs just to substitute again
$z^k \rightarrow \Omega_k (w) = z(w)^k_+-z(w)^k_-$ and
\be
\label{hflo}
z^k(\log z-c_k) \rightarrow
H_k(z,w) = z^k\log w + \sum_{j=1}^k C^{(k)}_j\Omega_j(w)
\ee
for the polynomial ${\bf t}$-flows and logarithmic ${\bf T}$-flows. The coefficients
$C^{(k)}_j$ in the r.h.s. of \rf{hflo} are completely fixed by the
asymptotic at $z\to\infty$ (see \cite{AMextT} for details).
The ${\bf T}$-dependence of $\F(a,{\bf t},{\bf T})$ is given by
\be
\label{TdF}
\left.{\d\F\over \d T_n}\right|_{\bf t} = (-)^{n} n!\left(S_n\right)_0
\ee
(inspired by K.~Saito formula \cite{Saito+}), where
\be
\label{Sn}
{d^nS_n\over dz^n} = S,\ \ \ n\geq 0
\ee
or $S_n$ is the $n$-th primitive of $S$ (odd under $w\leftrightarrow{1\over w}$). Certainly, the
particular $n=0$ case of the formula \rf{TdF} coincides with \rf{aDS0}, since the
variable $a$ corresponds to the primary $\sigma_0({\bf 1})\equiv {\bf 1}$ operator.
For $n=1$ formula \rf{TdF} gives rise to
\be
\label{Ft1T1}
\F (t_1,a,T_1) = {a^2t_1\over 2T_1} + T_1^2\exp{t_1\over T_1}
\ee
which at $T_1=1$ obviously coincides with \rf{fto}, while
at $T_1\to\infty$ gives
\be
\label{Ft1T1inf}
\F (t_1,a,T_1)\ \stackreb{T_1\to\infty}{\sim}\
\ldots {\sf F}_K(t_1+a,T_1) + {\sf F}_K(t_1-a,T_1) + \ldots
\ee
i.e. from what we started in \rf{utko}, \rf{fko} - a linear in $x$ solution to
the KdV equation $u(x,T_1) \sim {x\over T_1}$.

The ${\bf t}$-dependence of the quasiclassical tau-function $\F({\bf t},{\bf T})$ is
governed by dispersionless Toda chain hierarchy, while the ${\bf T}$-dependence can be encoded
in terms of so called extended Toda hierarchy \cite{EY,Dub}, which is rather special
nontrivial reduction \cite{Getz,OP} of the two-dimensional Toda lattice.
On the small phase space
\be
\label{eFsm}
\F_{\rm eq} (X_1,{\bar X}_1;\epsilon) = {\epsilon^2\over 6}\left(X_1^3+{\bar X}_1^3\right) + e^{X_1+{\bar X}_1}
= {a^2t_1\over 2}+e^{t_1}+{\epsilon\over 2}at_1^2+{\epsilon^2\over 6}t_1^3
\ee
coinciding with \rf{fto} at $\epsilon\to 0$, upon
\be
\label{tTx}
t_k = X_k + {\bar X}_k,\ \ \ \ k>0
\\
T_k = -\epsilon{\bar X}_{k+1},\ \ \ \ k\geq 0
\ee
and
\be
\label{gopder}
{\d\over\d X_k} = {\d\over\d t_k},\ \ \ {\d\over\d {\bar X}_k} = {\d\over\d t_k}
-\epsilon {\d\over\d T_{k-1}},\ \ \ k>0
\ee
for the derivatives,
indeed satisfies the two-dimensional Toda lattice equation
\be
\label{eToda}
{\d^2\F_{\rm eq}\over\d{\bar X}_1\d X_1} =
\exp{1\over\epsilon^2}\left({\d\over\d X_1}-{\d\over\d {\bar X}_1}\right)^2\F
\ee
if one takes the solutions, constrained by reduction
\be
\label{re2To}
{\d\F_{\rm eq}\over\d X_1}-{\d\F_{\rm eq}\over\d {\bar X}_1} = \epsilon{\d\F_{\rm eq}\over\d X_0}
\ee
On small phase space \rf{tTx} just give $X_1=t_1+{a\over\epsilon}$ and ${\bar X}_1=-{a\over\epsilon}$.

Equation of the curve for the equivariant Toda \cite{Getz}, has been derived recently
from the functional approach in \cite{AlNek}, can be written as
\be
\label{eqtocu}
z = v + \Lambda\left(w + {1\over w}\right) - \epsilon\log{z\over\Lambda w}
\ee
is a deformation of \rf{dtcu}, but it is already not algebraic. It should be also
supplemented by the conjugated one
\be
\label{eqbz}
{\bar z} = {\bar v} + \Lambda\left(w + {1\over w}\right) + \epsilon\log{{\bar z}w\over\Lambda}
\ee
with parameters $v$ and ${\bar v}$ obeying the reduction constraint \rf{re2To}, or
\be
\label{redv}
v-{\bar v} = \epsilon\log\Lambda^2
\ee
Equations \rf{eqtocu}, \rf{eqbz} can be solved hence for $z=z(w)$ and ${\bar z}={\bar z}(w^{-1})$ only
in terms of the expansions
\be
\label{zbzexp}
z\ \stackreb{w\to\infty}{=}\ \Lambda w + v +
\left(\Lambda-{\epsilon v\over\Lambda}\right){1\over w} + \ldots
\\
{\bar z}\ \stackreb{w\to 0}{=}\ {\Lambda\over w} + {\bar v} +
\left(\Lambda+{\epsilon {\bar v}\over\Lambda}\right)w + \ldots
\ee
producing, in particular, the two-dimensional Toda lattice Hamiltonians
$h_k=z^k(w)_+ + \half z^k(w)_0$ and
${\bar h}_k = {\bar z}^k(w^{-1})_- + \half {\bar z}^k(w^{-1})_0$, e.g.
\be
\label{tlham}
h_1(w) = \Lambda w + {v\over 2},\ \ \
{\bar h}_1(w^{-1}) = {\Lambda\over w} + {{\bar v}\over 2},
\\
h_2(w)= \Lambda^2w^2 + 2\Lambda vw + {v^2\over 2}
+ \Lambda^2-\epsilon v,\ \ \
{\bar h}_2(w^{-1}) =
{\Lambda^2\over w^2} + {2\Lambda {\bar v}\over w} + {{\bar v}^2\over 2}
+ \Lambda^2+\epsilon {\bar v}
\\
\ldots
\ee
Inverse to the first equation of \rf{zbzexp} has the form
\be
\label{invwz}
- \log {w\over z} = \log\Lambda + {v\over z} + \ldots
\ee
which means that
\be
\label{lvx}
\log\Lambda^2 = {\d^2\F_{\rm eq}\over\d X_0^2} = X_1 + {\bar X}_1
\\
v = {\d^2\F_{\rm eq}\over\d X_0\d X_1} = \epsilon X_1
\ee
and
\be
\label{bv}
{\bar v} = v - \epsilon\log\Lambda^2 = -\epsilon{\bar X}_1
= {\d^2\F_{\rm eq}\over\d X_0\d {\bar X}_1}
\ee
Solutions \rf{lvx} can be also found from the equations \rf{eqscu} for the
generating function
\be
\label{Seq}
S_{\rm eq} = \sum_{k>0}\left({Y_k\over k} h_k(w) - {{\bar Y}_k\over k} {\bar h}_k(w^{-1})\right) -
X_0\log w
\ee
where the ``exact'' Toda variables $\{ Y_k,{\bar Y}_k\}$, $k>0$ are related to the time variables
$\{X_k,{\bar X}_k\}$ \rf{tTx} by a triangular transformation \cite{Getz,OP}. In
particular
\be
\label{x1y1}
X_1=Y_1+1+{X_0\over\epsilon},\ \ \ {\bar X}_1={\bar Y}_1+1-{X_0\over\epsilon}
\ee
automatically giving rise to the constraint \rf{re2To}.

\section{Elliptic U(1) theory}

Before turning to the non Abelian gauge theories and corresponding target-space curves
of higher genera let us discuss the elliptic example of prepotential, being on one hand just a slight
generalization of the ``small phase space formula'' \rf{fto}, but on another hand which is quite
rare example of the case, when the prepotential can be computed explicitly beyond the
rational case. It corresponds literally to
the $U(1)$ gauge theory with massive adjoint matter, for which the target-space $\mathbb{P}^1$
should be replaced by elliptic curve, or complex torus with modulus $\tau$, playing the role
of $t_1$ or the ultraviolet coupling in gauge theory with vanishing beta-function.

Consider elliptic curve ${\cal E}_\tau = \mathbb{C}/(1,\tau)$ with a marked point $P$.
Choose $\xi(P)=0$ where $d\xi\in H^1({\cal E}_\tau)$ is the canonical holomorphic
differential
\be
\label{dzper}
\oint_A d\xi = 1,\ \ \ \ \ \oint_B d\xi = \tau
\ee
which appears here $d\Phi = {dw\over w}\rightarrow d\xi$ instead of the ``holomorphic''
third-kind Abelian differential on the cylinder \rf{dtcu}.
Consider also the second kind Abelian differential
\be
\label{dlam}
d\lambda = -Md\xi\left(\wp(\xi) + 2\eta\right) = -M{d\xi\over \xi^2} + \ldots
\ee
where the constant $\eta=\zeta(1/2)$ is chosen to ensure constancy of the periods
\be
\label{dlamper}
\oint_A d\lambda = 0
\\
\oint_B d\lambda = -M( 2\eta\tau - 2\eta') = - 2\pi i M = {\rm const}
\ee
with $\eta'=\zeta(\tau/2)$, and the second equation in the second formula
is due to the Legendre identity. Integrating \rf{dlam} one gets
\be
\label{lam}
\lambda = M(\zeta(\xi) - 2\eta \xi + i\pi) + a = M{\theta'(\xi)\over\theta(\xi)} + a + i\pi M
\ee
where $\theta(\xi) \equiv \theta_1(\xi|\tau)$ is the only odd Jacobi theta-function, while the
integration constant is taken to get
\be
\label{aperto}
\oint_A dS = \oint_A \lambda d\xi = a
\ee
For the dual $B$-period we have ($\xi_0$ is the intersection point of the $A$ and $B$ cycles)
\be
\label{bper}
\oint_B dS = \oint_B \lambda d\xi = (a + i\pi M)\tau + M\log{\theta(\xi_0+\tau)\over\theta(\xi_0)} =
\\
= a\tau + i\pi M(\pm 1 - 2\xi_0)
\ee
Taking derivatives of the generating one-form one gets
\be
\label{Sder}
{\d\over\d a} dS = \left.{\d\lambda\over\d a}\right|_{\xi,M,\tau}d\xi = d\xi
\\
{\d\over\d \tau} dS = \left.{\d\lambda\over\d \tau}\right|_{\xi,M,a}d\xi =
M\left({\d\over\d\tau}\log\theta(\xi)\right)'d\xi = 4\pi i M\left({\theta''(\xi)\over\theta(\xi)}\right)'d\xi
\ee
i.e. the derivative over $\tau$ gives rise to a non-single valued differential, since
\be
\label{omega}
\Omega = {\theta''(\xi)\over\theta(\xi)} = \left(\log\theta(\xi)\right)''
+ \left(\left(\log\theta(\xi)\right)'\right)^2 \equiv J'(\xi) + J^2(\xi)
\\
\Omega(\xi_0+1)-\Omega(\xi_0) = 0
\\
\Omega(\xi_0+\tau)-\Omega(\xi_0) = \left((J(\xi_0+\tau)-J(\xi_0)\right)\left((J(\xi_0+\tau)+J(\xi_0)\right)=
\\
= -4\pi i(J(\xi_0)-i\pi) = -4\pi i \left({\theta(\xi_0)'\over\theta(\xi_0)}-i\pi\right)
\ee
so that the jump (cf., for example, with \cite{KriW,BMMM})
\be
\label{jumpdom}
\Delta_A d\Omega = d\Omega(\xi_0+\tau)-d\Omega(\xi_0) =
-4\pi id\left({\theta(\xi_0)'\over\theta(\xi_0)}\right) = -{4\pi i\over M}d\lambda
\ee
In order to get the nontrivial part of prepotential one needs now to compute
\be
\label{Ftau}
{\d\F\over\d\tau} = \ha\oint_A \lambda^2 d\xi = \ha \left(a^2 + \pi^2M^2 +
M^2\oint_A \left({\theta(\xi)'\over\theta(\xi)}\right)^2 d\xi\right)
\ee
For the last integral one gets
\be
\label{serint}
\oint_A \left({\theta(\xi)'\over\theta(\xi)}\right)^2 d\xi = \pi^2\int_{\xi_0}^{\xi_0+1} \cot^2(\pi \xi) d\xi
+ 8\pi^2\sum_{n>0}{q^{2n}\over 1-q^{2n}}\int_{\xi_0}^{\xi_0+1} \cot(\pi \xi)\sin(2\pi n\xi) d\xi +
\\
+ 16\pi^2\sum_{n,k>0}{q^{2n}\over 1-q^{2n}}{q^{2k}\over 1-q^{2k}}
\int_{\xi_0}^{\xi_0+1} \sin(2\pi n\xi)\sin(2\pi k\xi) d\xi
\ee
where $q=e^{i\pi\tau}$, and
\be
\label{intcomp}
\int_{\xi_0}^{\xi_0+1} \cot^2(\pi \xi) d\xi = - \pi^2
\\
\int_{\xi_0}^{\xi_0+1} \cot(\pi \xi)\sin(2\pi n\xi) d\xi = 1,\ \ \ \ n>0
\\
\int_{\xi_0}^{\xi_0+1} \sin(2\pi n\xi)\sin(2\pi k\xi) d\xi = \half \delta_{n,k},\ \ \ n,k>0
\ee
Therefore
\be
\label{Ftaures}
{\d\F\over\d\tau} = \ha\oint_A \lambda^2 d\xi =
{a^2\over 2} +4\pi^2 M^2\sum_{n>0}{q^{2n}\over (1-q^{2n})^2}
\ee
Integrating this relation one finally gets
\be
\F = {\tau a^2\over 2} + {2\pi M^2\over i}\sum_{n>0}\int {q^{2n-1}dq\over (1-q^{2n})^2}
= {\tau a^2\over 2} + {\pi M^2\over i}\sum_{n>0}{1\over n}\left({1\over 1-q^{2n}}-1\right)
\ee
Ressumation of the last expression gives
\be
\sum_{n>0}{1\over n}\left({1\over 1-q^{2n}}-1\right) =  \sum_{n,k>0}{q^{2nk}\over n}
= - \sum_{k>0}\log\left( 1-q^{2k}\right)
\ee
so that
\be
\label{prep}
\F = \half \tau a^2 - m^2\log\prod_{k>0}\left( 1-q^{2k}\right) \sim
\half \tau a^2 - m^2\log\eta(\tau)
\ee
where the ``physical mass'' is $m^2={\pi\over i} M^2$, and
$\eta(\tau) = q^{1/12}\prod_{k>0}\left( 1-q^{2k}\right)$ is the Dedekind function
(extracting from or adding to prepotential linear term in $\log q\sim\tau$ is
inessential). In the limit $m\to\infty$, $\tau\to +i\infty$ with
\be
\label{inoli}
m^2e^{2\pi i\tau} = \Lambda^2 = e^{t_1} = {\rm fixed}
\ee
one comes back from \rf{prep} to \rf{fto} (again, generally up to inessential maximally quadratic
terms). It is quite unusual fact, that the Dedekind function $\eta(q)$ appears in the
main quasiclassical or $g=0$ term in the genus expansion for the generating function
$\log\tau = {1\over \hbar^2}\F + F_1 + \hbar^2 F_2 + \ldots$, while commonly
one can rather expect its appearance in the $F_1 \sim \log\det{\bar\d}$. This can be
related with the nontrivial questions about the modular properties of $F_g$ and the
holomorphic anomaly equations \cite{BCOV,DKl,LMN}, which are beyond the scope of these notes.

\section{Non Abelian theory}

Topological gauge/string duality claims in particular, that the truncated Gromov-Witten genus zero
potential $\F(a,{\bf t})$ coincides with a particular oversimplified $N_c=1$ case of the
(extended) Seiberg-Witten
\N2 supersymmetric gauge theory. Moreover, it turns out that quasiclassical solution to the non Abelian
theory with the gauge group $U(N_c)$ can be obtained almost in the same way \cite{MN} - extremizing the
functional \rf{functnl}, with the only slight modification of the constraint part of the problem.
Instead of a single constraint $\half\int dx\ xf''(x)=a$ one has now to impose a set of $N_c$
similar constraints
\be
\label{noaco}
\half\int_{{\bf I}_i} dx\ xf''(x)=a_i,\ \ \ i=1,\ldots,N_c
\ee
for the shape function $f(x)$, corresponding to $N_c$-tuples of partitions \cite{NO}, with
the vertices
at $a_1,\ldots,a_{N_c}$ correspondingly. All these constraints are taken into account, just
as in \rf{functnl}, by adding them to the functional with the Lagrange multipliers
$a^D_1,\ldots,a^D_{N_c}$.

To solve the extremal equations under the set of new constraints, one now takes the double
cover of $z$-plane with $N_c$ cuts $\{ {\bf I}_j: x^-_j<z<x^+_j \}$
\be
\label{dcN}
y^2 = \prod_{i=1}^{N_c}(z-x^+_i)(z-x^-_i)
\ee
or the hyperelliptic curve of genus $N_c-1$, and constructs $S$, odd under the involution
$y \leftrightarrow -y$ or pure imaginary on the set ${\bf I}$ of these $N_c$ cuts
${\bf I}=\cup_{j=1}^{N_c}{\bf I}_j$.

Such non Abelian extended $U(N_c)$ \N2 supersymmetric gauge theory is solved now
in terms of the Abelian
differentials. The functional equation \rf{Scut}
is solved by constructing the differential of
\be
\label{dS}
\Phi(z) = {dS\over dz}
=\sum_{k>0}kt_k z^{k-1}-\half\int dx f''(x)\log(z-x)
\ee
(remember that the shape function is
restored from the jump $f'(x) \sim {\rm jump}\ \Phi(x)$). On
hyperelliptic curve \rf{dcN}, one writes for $d\Phi$
\be
\label{dpsiy}
d\Phi = \pm\ {\phi(z)dz\over y} = \pm\ {\phi(z)dz\over\sqrt{\prod_{i=1}^{N_c}(z-x^+_i)(z-x^-_i)}}
\ee
where the numerator $\phi(z)$ is totally fixed by asymptotic and the periods
\be
\label{dpsiperA}
\oint_{A_j} d\Phi = -2\pi i \int_{{\bf I}_j} f''(x) dx =
-2\pi i \left(f'(x_j^+)-f'(x_j^-)\right) = -4\pi i
\ee
If all $t_k=0$, for $k>1$, $t_1=\log\Lambda^{N_c}$ (of course still $T_n=\delta_{n,1}$!)
at the vicinity of the points $P_\pm$, where $z(P_\pm)=\infty$, one finds
\be
\label{fiinf}
\Phi\
\stackreb{P\to P_\pm}{=}\  \mp 2N_c\log z \pm 2N_c\log\Lambda  + O(z^{-1})
\ee
and it means, that there exists a meromorphic function
$w=\Lambda^{N_c}\exp\left(-\Phi \right)$, satisfying equation \rf{Todacu}.
At $N_c=1$ we come back to the curve \rf{dtcu}, or the Lax operator of dispersionless Toda chain.

If however the higher $t_k\neq 0$ are nonvanishing, $\exp\left(-\Phi \right)$ has an essential
singularity and cannot be described algebraically. Implicitly everything is still fixed by the
asymptotic
\be
\label{dfas}
d\Phi\ \stackreb{z\to\infty}{\sim}\ \sum_{k>1} k(k-1)t_kz^{k-2}+\ldots
\ee
and the period constraints \rf{dpsiperA}, implying in particular
\be
\label{delShol}
\delta (dS) = \delta \left(\Phi dz\right)\
\stackreb{z\to x^\pm_j}{=}\ {-\phi(x^\pm_j)\delta x^\pm_j \over\prod'_k
\sqrt{(x^\pm_j-x^+_k)(x^\pm_j-x^-_k)}}{dz\over
\sqrt{z-x^\pm_j}}+\ldots
\\
\simeq {\rm holomorphic}
\ee
the generic analog of the conditions \rf{eqscu}.

The A-period constraints together with the asymptotic \rf{dfas} fix completely the form
of the differential $d\Phi$. Vanishing of the B-periods together with the integrality
of the coefficient in front of the logarithm in \rf{fiinf}
impose $N_c$ constraints for the $2N_c$ ramification points of the curve \rf{dcN}.
Remaining $N_c$ yet unknowns are ``eaten'' by the Seiberg-Witten periods
\be
\label{SWper}
a_i =  {1\over 4\pi i}\oint_{A_i} z d\Phi,\ \ \ i=1,\ldots,N_c
\ee
where the extra $N_c$-th dependent period is also included, an alternative option is to fix the
residue at infinity $a$, as in \rf{ares}. The dual to \rf{SWper} B-periods define the
gradients of the prepotential
\be
\label{duper}
a^D_i = \half\oint_{B_i} zd\Phi  = {\d\F\over\d a_i},\ \ \ i=1,\ldots,N_c
\ee
and, as in the $U(1)$ case, the ${\bf t}$-derivatives are determined by the
residue formulas
\be\label{Ftres}
{\d \F \over \d t_k} =
- {1\over k+1}{\rm res}_{P_+} \left( z^{k+1}d\Phi \right),\ \ \ k>0
\ee
but now on the curve \rf{dcN}.
Integrability condition \rf{prepsw} for the gradients \rf{duper}
is ensured by the symmetricity of the period matrix of \rf{dcN}, or more generally, when
including formulas \rf{Ftres}, -
by the Riemann bilinear identities for all Abelian differentials.

If higher $T_n\neq 0$ are nonvanishing, say we consider first $N$ descendants of unity
to be switched on for $n=1,\ldots,N$
\footnote{The ``minimal'' theory has $T_n=\delta_{n,1}$ and
$\F=\F(a,{\bf t})$; $T_1=1$ corresponds to the nonvanishing
condensate $\langle\sigma_1(\varpi)\rangle\neq 0$.},
only the $(N+1)$-th derivative of \rf{Szfu}
\be
\label{fin1}
d\Phi^{(N-1)} = d\left({d^N S\over dz^N}\right)
\ee
can be decomposed over the basis of single-valued on the curve \rf{dcN}
Abelian differentials.
It is again totally determined by integrality of A-periods and singularities, but now not
only at $z(P_\pm)=\infty$, but also at the branch points $\{ x_j\}$,
$j=1,\ldots,2N_c$ of the curve \rf{dcN}, where the differential \rf{fin1} acquires extra poles.
In fact these extra singularities are artificial, and $\Phi',\ldots,\Phi^{(N-1)}$ are regular
at the branching points, if being considered as
$2-,\ldots,N-$ differentials on the curve \rf{dcN}.

To construct \rf{fin1} explicitly one writes
\be
\label{fin1hy}
d\Phi^{(N-1)} = {\phi(z)dz\over y} + {dz\over y}\sum_{j=1}^{2N_c}\sum_{k=1}^{N-1}
\left({q^k_j\over(z-x_j)^k}\right)
\ee
and fixes the periods of $d\Phi^{(N-1)},d\Phi^{(N-2)},\ldots,d\Phi',d\Phi$
by $2N_c\cdot N$ constraints, ending up, therefore with
\be
\label{count}
(2N+1)N_c - 2N_c\cdot N = N_c
\ee
variables, to be absorbed by
the generalized Seiberg-Witten periods
\be
\label{SWperA}
a_j = {1\over 4\pi i}\oint_{A_j} {z^N\over N!}d\Phi^{(N-1)},\ \ \ \ j=1,\ldots,N_c
\ee
which define the prepotential still by
\be
a^D_j = \half\oint_{B_j} {z^N\over N!}d\Phi^{(N-1)} = {\d\F\over\d a_j},\ \ \ \ j=1,\ldots,N_c
\ee
The generalized Seiberg-Witten form is now the $N$-tuple Legendre transform of $S$-function \rf{Szfu},
being certainly a multivalued Abelian integral on the curve \rf{dcN}.

\section{Two functional formulations}

In the perturbative limit $\Lambda\to 0$ the cuts of the curve \rf{dcN} shrink to the points
$z=v_j$, $j=1,\ldots,N_c$ and the curve becomes rational, possibly parameterized as
\be
w_{\rm pert} = P_{N_c}(z) = \prod_{i=1}^{N_c} (z-v_i)
\ee
This curve is endowed now with two polynomials (of arbitrary power):
${\bf t}(z)\equiv\sum_{k>0}t_kz^k$ and $T(x)\equiv\sum_{n>0}T_nx^n$. The $S$-functions
is computed in this case explicitly and reads
\be
\label{SFiN}
S(z) = {\bf t}'(z) -2\sum_{j=1}^{N_c} \sigma(z;v_j) = {\bf t}'(z) -
2\sum_{j=1}^{N_c}\sum_{k>0}{T^{(k)}(v_j)\over k!}(z-v_j)^k(\log(z-v_j)-c_k)
\ee
i.e. is defined in terms of the function
\be
\label{sifun}
\sigma(z;x) = \sum_{k>0}{T^{(k)}(x)\over k!}(z-x)^k(\log(z-x)-c_k)
\ee
where the sum is finite, if restricted to the $N$-th class of backgrounds,
with only first times $T_1,\ldots,T_N \neq 0$.

The perturbative prepotential is defined by the gradients
\be
\label{Sprepert}
a^D_i = S(v_i) = {\d\F_{\rm pert}\over\d a_i}
\ee
and this formula gives rise to the following explicit expression
\be
\label{pertprep}
\F_{\rm pert} (a_1,\ldots,a_{N_c};{\bf t},\textbf{T}) =
\sum_{j=1}^{N_c} F_{UV}(a_j;{\bf t},\textbf{T})+
 \sum_{i\neq j}F(a_i,a_j;\textbf{T})
\\
a_j = T(v_j),\ \ \ \ j=1,\ldots,{N_c}
\ee
with
\be
\label{cufuv}
F_{UV}(x) \equiv F_{UV}(x;{\bf t},\textbf{T}) = \int_0^x {\bf t}'({\sf x})dT({\sf x})
\\
{\d^2\over\d x_1\d x_2}F(x_1,x_2;\textbf{T}) = T'(x_1)T'(x_2)\log(x_1-x_2)
\ee
If $T_n=\delta_{n,1}$ one gets from \rf{cufuv}
\be
\label{fuvtru}
\left.F_{UV}(x)\right|_{T_n=\delta_{n,1}} = \sum_{k>0}t_k{x^{k+1}\over k+1}
\ee
which is the ultraviolet prepotential for the switched off descendants of unity, except
for the condensate of $\langle\sigma_1({\bf 1})\rangle$, and
\be
\label{ketr}
\left.F(x_1,x_2;\textbf{T})\right|_{T_n=\delta_{n,1}} = F(x_1-x_2)
\ee
where the r.h.s. depends only upon the difference of the arguments, is ${\bf T}$-independent
and reproduces the constant kernel from the r.h.s. of \rf{gamf}. Generally, switching
on the descendants of unity one induces a reparameterization in the moduli space from $a=v$ to
$a=T(v) = \sum_n T_nv^n$ which also results in crucial complifications in the
formulas \rf{cufuv}.

The perturbative prepotential \rf{pertprep} defines therefore the general
form of the functional for ${{\cal F}} ( {\bf a}, {\bf t}, {\bf T} )$ given by
\be
\label{functgen}
\F = -\half\int_{x_1>x_2} dx_1 dx_2 f''(x_1)f''(x_2) F(x_1,x_2;{\bf T}) +
\int dx f''(x)F_{UV}(x;{\bf t},{\bf T}) + \\
+ \sum_i a^D_i\left(a_i-\half\int dx\ xf''(x)\right)
+ \sigma\left(1-\half\int dx\ f''(x)\right)
\ee
The functional \rf{functgen} can be treated, except for one point we specially pay attention to
below, just in the same way as the functional \rf{functnl} at $T_n=\delta_{n,1}$. For example,
since the kernel (the second formula from \rf{cufuv}) is ${\bf t}$-independent, one finds
that
\be
\label{tderg}
{\d\F\over \d t_k} = \int dx f''(x){\d F_{UV}(x;{\bf t},\textbf{T})\over \d t_k}
\ee
are still given by the ``regular tail'' of the expansion of
\be
\label{Ssigma}
S(z) = {d\over dz}{\delta\F\over\delta f''(z)} =  {\bf t}'(z) - a^D - \int dxf''(x) \sigma(z;x)
\ee
However, a problem with the functional \rf{functgen} is in computing
the derivatives ${\d\F\over \d T_n}$, since the kernel $F(x_1,x_2;\textbf{T})$
is ${\bf T}$-dependent. Fortunately, there exists an equivalent
alternative functional formulation, related to \rf{functgen} by an integral transform
$\int dx\rho(x)g(x) = \int dx f''(x){\hat D}_{N-1}(x)g(x)$, so that
\be
\label{fuN}
\F = \F_N\left[\rho\right] =
{1\over 2}\int dx \rho(x){\bf t}_N(x)
+ {(-)^{N}\over (2N)!}\int_{dx_1 dx_2}  \rho(x_1)\rho(x_2) H_{2N}^{(+)}(x_1-x_2) +
\\
+\sum_{n=0}^N\sigma_n\left(T_n-{(-)^{n-1}\over 2}\int dx {x^{N-n}\over (N-n)!}\rho(x)\right)
\ee
where ${\bf t}_N(x) = \sum_{k>0} t_k {x^{k+N}\over (k+1)\ldots(k+N)}$, the kernel
${(-)^{N-1}\over (2N)!}H_{2N}^{(+)}(x) = {1\over (2N)!}x^{2N}\left( \log x - c_{2N}\right)$
is ${\bf T}$-independent, and
\be
{\hat D}_{N-1}(x) = T'(x){d^{N-1}\over dx^{N-1}}-T''(x){d^{N-2}\over dx^{N-2}}+\ldots
+(-)^{N-1}T^{(N)}
\ee
is the $(N-1)$-th order differential operator.

A nontrivial consequence of the alternative functional formulation \rf{fuN}
is the multiple-primitives Saito formula \rf{TdF}
\be
\label{sigmasem}
{\d\F\over\d T_n} = \sigma_n = (-)^{n} n!\left(S_n\right)_0, \ \ \ n=0,\ldots,N
\ee
where the right equality expresses the Lagrange multipliers from the formula \rf{fuN} in terms of the
constant parts of
\be
\label{derexN}
S_{N-1}(z) = \textbf{t}_{N-1}(z) - {(-)^{N-1}\over (2N-1)!}\int dx \rho(x) H_{2N-1}^{(+)}(z-x) +
\sum_{n=0}^{N-1} \sigma_n (-)^{n}{z^{N-n-1}\over (N-n-1)!}
\\
\vdots
\\
S(z) =\textbf{ t}'(z) - {(-)^{N-1}\over N!}\int dx \rho(x) H_{N}^{(+)}(z-x) + \sigma_0
\ee
a sequence of functions, whose real parts vanish on the set of the cuts.

\section{Conclusion}

It has been demonstrated above, that
instead of solving complicated problems of quantum field theory or
string theory directly
one can sometimes solve instead the simple differential equations. The
desired solutions for the purposes of topological string and gauge theories
are very strange from conventional point of view. Nevertheless, they are still
related to geometry of complex curves, and therefore in many cases - in contrast with
the case of higher-dimensional complex manifolds - can be described and calculated explicitly.

The simplified $N_c=1$ extended Seiberg-Witten theory is dual to the topological type A string, or the
Gromov-Witten theory of projective line $\mathbb{P}^1$. Such Abelian $N_c=1$ theory is solved
completely in terms of dispersionless Toda chain hierarchy, and this solution is
specified by unavoidable presence of the topological Eguchi-Yang term. Generally, for the
full Gromov-Witten potential one gets the solution to the so called extended Toda chain hierarchy -
extended by the flows, corresponding to the switched on descendants of the unity operator.
Extended chain hierarchy can be also seen as a limiting case of the so called equivariant Toda lattice -
a very special reduction of the two-dimensional Toda lattice hierarchy. We have presented above, how
the geometry of this reduction on the small phase space looks like explicitly.

To extend these results to the
non Abelian gauge theories one needs, however, to use the technique of constructing quasiclassical
tau-functions in terms of Abelian integrals on hyperelliptic
curve of arbitrary genus $N_c-1$. We have demonstrated, how it can be performed explicitly
on the elliptic curve, giving rise to the straightforward generalization of
the Toda chain solution on small phase space in terms of the modular Dedekind function,
and presented the generic implicit formulation for
higher genera. It is important to point out, that extension of the Seiberg-Witten theory,
similar to switching on all gravitational descendants, requires extension of the basis
of Abelian differentials, which should contain necessarily the differentials with extra points
in the branching points of the curve. We have also discussed the delicate issues of the
functional formulation in the extended case, and proposed an integral transform, relating two
different forms of the functional, useful for getting answers to different questions.

{\bf Acknowledgements}.
I am grateful to A.~Alexandrov, M.~Kazarian, I.~Krichever
and, especially, to N.~Nekrasov for the very useful discussions.
The work was partially supported by the Federal Nuclear Energy Agency, the RFBR grant
08-01-00667,
the grant for support of Scientific Schools LSS-1615.2008.2,
the INTAS grant 05-1000008-7865, the project ANR-05-BLAN-0029-01, the
NWO-RFBR program 047.017.2004.015, the Russian-Italian RFBR program 06-01-92059-CE, and by the
Dynasty foundation.

\end{document}